\documentclass[a4paper,11pt]{article}
\usepackage{pos}
\usepackage[separate-uncertainty=true]{siunitx}
\usepackage{lineno}

\usepackage{romannum}
\AtBeginDocument{\pagenumbering{arabic}}

\usepackage{enumitem}
\usepackage{wrapfig}

\usepackage{xspace}

\newcommand{\PAO}{Pierre Auger Observatory\xspace}
\newcommand{\NXIX}{\ensuremath{N_\text{19}}\xspace}
\newcommand{\Srad}{\ensuremath{S_\text{rad}}\xspace}

\newcommand{\Eem}{\ensuremath{E_\text{EM}}\xspace}
\newcommand{\Xmax}{\ensuremath{X_\text{max}}\xspace}

\title{Recent Highlights from the Auger Engineering Radio Array}

\author*[a]{Marvin Gottowik}
\affiliation[a]{Karlsruhe Institute of Technology, Institute for Astroparticle Physics, Karlsruhe, Germany}

\onbehalf{for the Pierre Auger Collaboration$^c$}
\affiliation[c]{Observatorio Pierre Auger, Av.\ San Mart{\'\i}n Norte 304, 5613 Malarg\"ue, Argentina\\
Full author list: {\rm\url{https://www.auger.org/archive/authors_icrc_2025.html}}}

\emailAdd{spokespersons@auger.org}

\abstract{
The Auger Engineering Radio Array (AERA) consists of 153 autonomous antenna stations deployed over 17 km² to measure the radio emission from extensive air showers initiated by cosmic rays with energies between 0.1 and 10 EeV in the 30 to 80 MHz frequency band. It operates in coincidence with the other detectors of the Pierre Auger Observatory particularly the Surface Detector (SD) and the Fluorescence Detector (FD). As the largest cosmic-ray radio detector worldwide before the recent Observatory upgrade AugerPrime, AERA has played a pioneering role in the development of radio technique for cosmic rays, providing complementary measurements and serving as a testbed for ideas that motivated the build-up of a new Radio Detector (RD) as part of AugerPrime. We report on measurements of the depth of the shower maximum using the radio footprint, demonstrating compatibility and competitive resolution with established FD results. An absolute calibration of AERA is achieved by monitoring the sidereal modulation of the diffuse Galactic radio emission for nearly a decade, confirming the long-term stability of a radio detector with no significant aging effects observed. This stability suggests that radio detectors could also be used to monitor potential aging effects in other detector systems. Additionally, we investigate the muon content of inclined air showers using hybrid SD-AERA events. Our results indicate that the muon content in measured data is consistent with expectations for iron nuclei as predicted by current-generation hadronic interaction models, confirming the well-known muon deficit for the first time with radio data. These findings reinforce the value of radio detection for cosmic-ray studies and provide a foundation for the next generation of analyses with the AugerPrime RD.
}

\ConferenceLogo{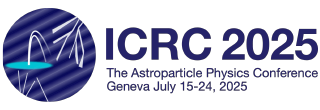}
\FullConference{39th International Cosmic Ray Conference (ICRC2025)\\
 15–24 July 2025\\
Geneva, Switzerland\\}

\begin{document}
\maketitle

\section{Introduction}

Over the past decade, radio detection has become a powerful technique for the study of extensive air showers, offering complementary insights alongside established optical and particle-based detection methods~\cite{Huege:2016veh}. Since the radio emission is produced exclusively by the electromagnetic component of the shower, its detection provides information that complements measurements from other detectors. The radio technique offers key advantages over the fluorescence and air-Cherenkov methods due to its almost \SI{100}{\percent} duty cycle. It is only limited by strong atmospheric electric fields that occur during thunderstorms and in the presence of large rain clouds. In addition, the radio signal is not attenuated as the atmosphere is transparent to radiation in the MHz range. The radio emission contains information about the arrival direction and energy of the primary cosmic ray, and is also sensitive to the longitudinal development of the air shower, particularly the depth of the shower maximum, \Xmax, which is statistically correlated with the mass of the primary particle.

The Auger Engineering Radio Array (AERA), as part of the Pierre Auger Observatory~\cite{auger_nim}, has played a central role in this development. Designed to probe the radio emission from air showers in the \SIrange{30}{80}{MHZ} band, AERA has provided critical experimental benchmarks for cosmic rays above \SI{e17}{eV} up to \SI{e19}{eV}. In this contribution, we present a selection of recent highlights from AERA. These include measurements of the depth of the shower maximum~\cite{AERA_Xmax_PRD, AERA_Xmax_PRL}, long-term absolute calibration using Galactic background emission~\cite{galactic_calib}, and novel insights into the muon content of inclined showers~\cite{muon_content}. An independent analysis of the absolute energy scale with AERA is presented in Ref~\cite{AERA_energy_scale}. Collectively, these results underline the robustness and scientific potential of radio detection in the ultra-high-energy regime.

\section{The Auger Engineering Radio Array (AERA)}

\begin{figure}
    \centering
    \includegraphics[width=0.95\linewidth]{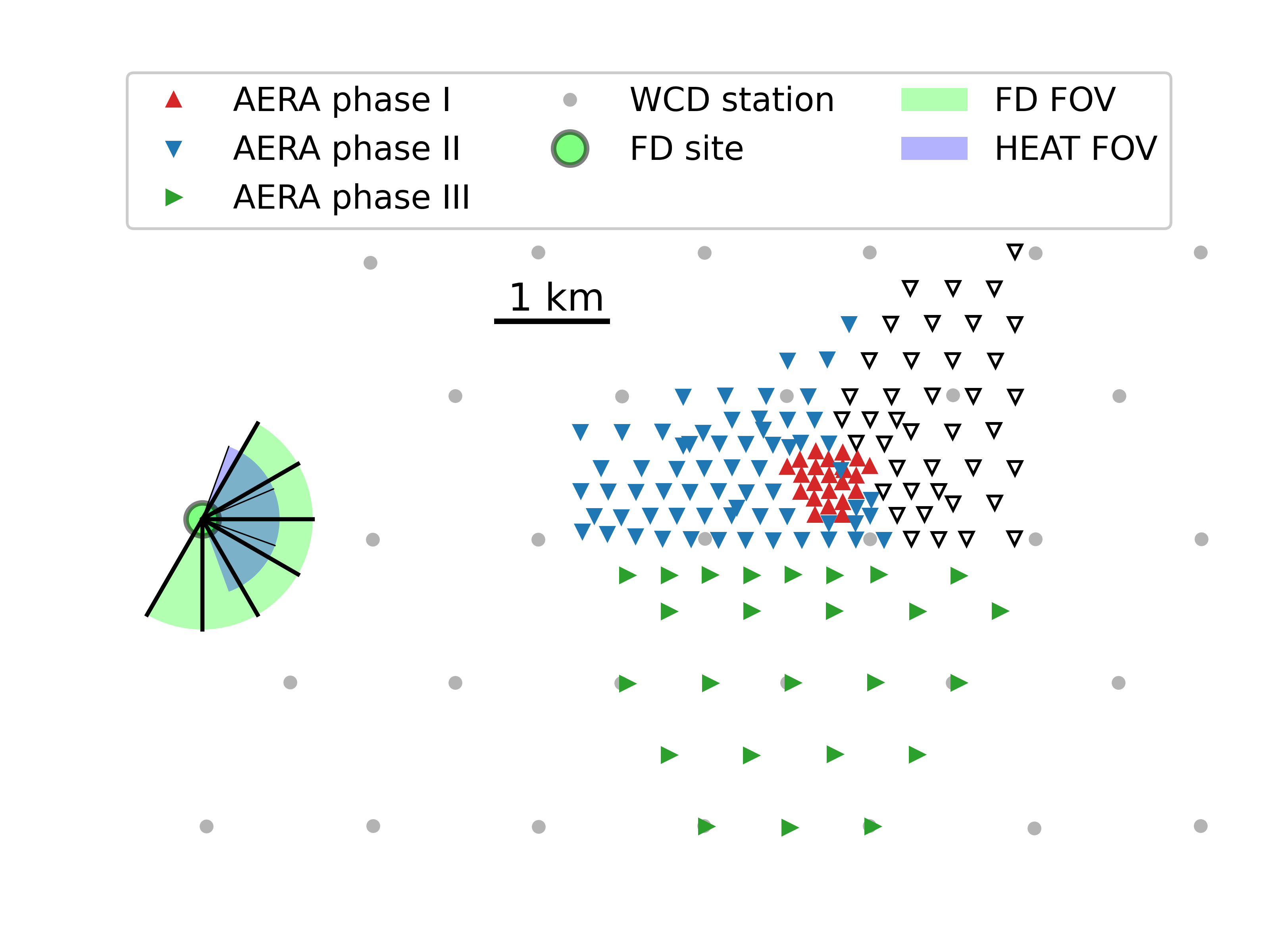}
    \caption{Map of AERA. The orientations of the triangles indicate the three deployment phases, empty triangles represent stations that cannot receive an external trigger. Only the water-Cherenkov stations with a grid spacing of 1500 m are shown.}
    \label{fig:AERA-map}
\end{figure}

AERA is located in the northwestern part of the Surface Detector (SD). With its 153 autonomous radio stations deployed over \SI{17}{km^2}, AERA was the largest operational radio array for cosmic rays until the advent of the AugerPrime Radio Detector (RD)~\cite{RD}. It was deployed in three phases, gradually increasing its coverage and station spacing. The first 24 stations (AERA phase \Romannum{1}) were installed in 2011 on a \SI{144}{m} triangular grid, covering \SI{0.4}{km^2}. In 2013, an additional 100 stations (AERA phase \Romannum{2}) were deployed with a larger spacing of \SI{250}{m} to \SI{375}{m}, expanding the array to \SI{6}{km^2}. Finally, in 2015, the last 29 stations (AERA phase \Romannum{3}) were added with a spacing of up to \SI{750}{m}, completing the current layout. A map of the individual deployment phases of AERA is presented in Fig.~\ref{fig:AERA-map}. Logarithmic periodic dipole antenna (LPDAs) were deployed in phase \Romannum{1} and Butterfly antennas were used in phases \Romannum{2} and \Romannum{3}~\cite{JINST2012P10011}.

AERA has been operating continuously since 2011 in coincidence with both the SD~\cite{ALLEKOTTE2008409} and Fluorescence Detector (FD)~\cite{ABRAHAM2010227}, providing a stable dataset for long-term studies of air showers and detector performance. The SD comprises more than 1600 water Cherenkov detector stations distributed across \SI{3000}{km^2} arranged in three nested triangular grids with spacings of \SI{1500}{m} (SD1500), \SI{750}{m} (SD750), and \SI{433}{m} (SD433). It provides information on the lateral distribution of the particles in the air showers on ground. The FD consists of 27 telescopes at four sites that observe the faint ultraviolet light emitted by atmospheric nitrogen excited by shower particles, enabling reconstruction of the longitudinal profile and the depth of shower maximum.

Its hybrid operation  has enabled measurements of air shower properties with high precision, combining complementary information on the electromagnetic and muonic shower component, and the longitudinal shower development. Furthermore, AERA now serves as a testbed for the development of advanced reconstruction techniques and calibration methods, many of which will be transferred to the RD. 

\section{Measurements of the depth of air-shower maximum}

The depth of shower maximum, \Xmax, is a key observable for determining the mass composition of cosmic rays. \Xmax can be inferred from the shape and structure of the radio footprint on the ground, which is sensitive to the geometric distance between the emission region and the antenna array. As the location of the bulk of the radio emission is associated with shower maximum, the resulting footprint encodes information about it.

To reconstruct \Xmax, a set of detailed air-shower simulations is performed with CoREAS~\cite{Huege:2013vt} for each measured high-quality event using proton and iron nuclei as primary particles to cover a wide range of \Xmax values. We focus on events arriving from within \ang{55} of the zenith because sensitivity to \Xmax decreases for higher inclinations as it will be more distant. These simulations include realistic atmospheric conditions from GDAS data, a time-dependent geomagnetic field, and measured background noise to accurately reproduce the detector environment. The simulated radio footprints are then compared with the measurements using a likelihood-based approach, which allows identifying the most probable \Xmax, assessing reconstruction biases, and determining resolution.

\begin{figure}
    \centering
    \includegraphics[width=\linewidth]{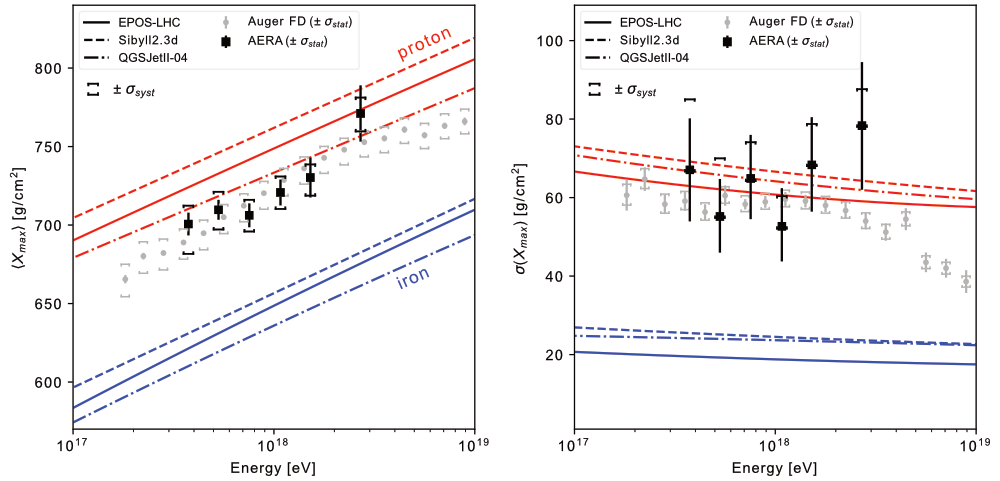}
    \caption{First two moments of the \Xmax distribution as measured by AERA as a function of SD-measured cosmic-ray energy. Model lines for proton and iron nuclei are added for comparison. Also, the FD-measured moments are shown for comparison. The  vertical bars indicate statistical uncertainties and the capped markers show the systematic uncertainties.}
    \label{fig:Xmax_momente}
\end{figure}

We validate the method by comparing the reconstructed \Xmax values to those obtained with the FD on an event-by-event basis for a subset of 53 high-quality hybrid events, where both AERA and the FD provide independent reconstructions. This comparison shows no significant bias between the two methods. The resolution and composition-sensitive observables are then evaluated using a larger dataset of approximately 600 radio-only events. The \Xmax resolution improves with the energy of the primary particle and reaches values below \SI{15}{g/cm^2}, comparable to the FD. Furthermore, the distribution of \Xmax values reconstructed from AERA data is compatible with FD measurements, as shown in Fig.~\ref{fig:Xmax_momente}, both in terms of the first two moments and the overall distribution shapes.

\section{Muon content of inclined air showers}

The number of muons in extensive air showers is a key observable in the study of ultra-high-energy cosmic rays. It carries complementary information on the mass of the primary particle and the hadronic interactions that govern the early stages of the shower development. A persistent challenge in this context has been the so-called ``muon deficit'', the observation that air shower simulations systematically underestimate the number of muons measured at ground level. This discrepancy has been observed by previous studies conducted at the \PAO~\cite{2015PhRvD..91c2003A, PierreAuger:2021qsd, Aab:2020frk}, while other experiments, such as Yakutsk, do not report a significant discrepancy~\cite{Glushkov2023}. A broader overview of results from nine air-shower experiments is given in Ref~\cite{Soldin:2021wyv}, highlighting the need for further investigation and novel approaches. In the following, we present results from a novel and independent approach to measure the muon content with hybrid events detected by the SD station deployed on a \SI{1500}{m} grid and AERA in coincidence.

For inclined air showers with zenith angles greater than \ang{60}, the electromagnetic component of the air shower is largely absorbed in the atmosphere and predominantly muons are detected by particle detectors on the ground. However, the radio emission arising from the electromagnetic component of the air shower is well understood and unaffected by atmospheric absorption or scattering making it a robust tool for energy estimation~\cite{PhysRevLett.116.241101, PierreAuger:2015hbf}.

For the SD reconstruction, we use the well-established method for inclined showers described in Ref~\cite{PierreAuger:2014jss}, which is fully efficient for primary energies above \SI{4}{EeV}. To good approximation, the shape of the muon distribution on the ground is found to be independent of the primary particle type, energy, and hadronic interaction model used in simulations. Therefore, reference maps of the muon density distribution on the ground as predicted by QGSJet II-03~\cite{PhysRevD.74.014026} at an energy of \SI{e19}{eV} are rescaled in the reconstruction to match the measured signals of the SD stations, i.e.
\begin{equation}
    \label{eq:NXIX}
    \rho_\mu(\vec{r};\theta,\phi,E) = \NXIX \, \rho_{\mu,19}(\vec{r};\theta,\phi).
\end{equation}
The rescaling factor, \NXIX, serves as a relative measure of the number of muons compared to the reference model and can also be used as an energy estimator of the cosmic ray.

For the radio signal, the signal distribution on the ground is described with a model specifically designed for inclined air showers with zenith angles above \ang{65}~\cite{Schlueter_2023}. As the emission is beamed, integrating the lateral distribution function (LDF) over the whole footprint yields the radiation energy. Applying corrections for air density and geomagnetic angle, we obtain the ``corrected radiation energy'', \Srad, which is directly related to the energy of the electromagnetic particle cascade, \Eem $\propto$ $\sqrt\Srad$~\cite{Glaser_2016}.

\begin{figure}
    \centering
    \includegraphics[width=.5\linewidth]{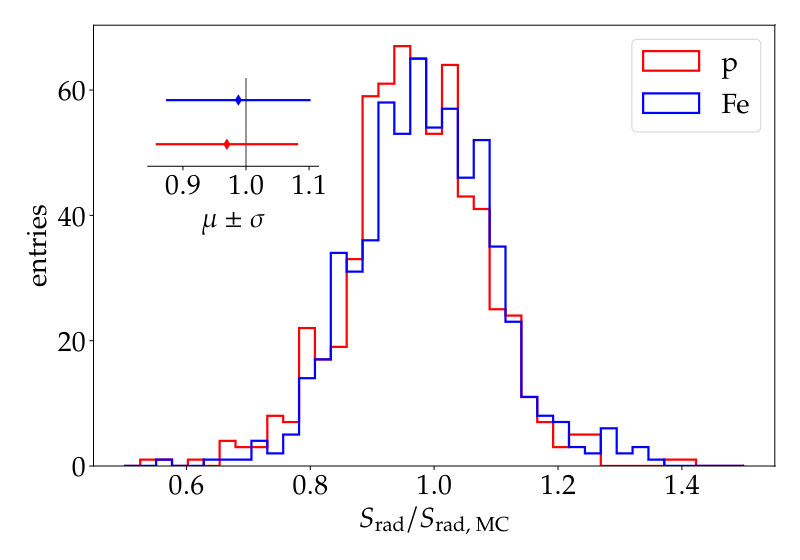}%
    \includegraphics[width=.5\linewidth]{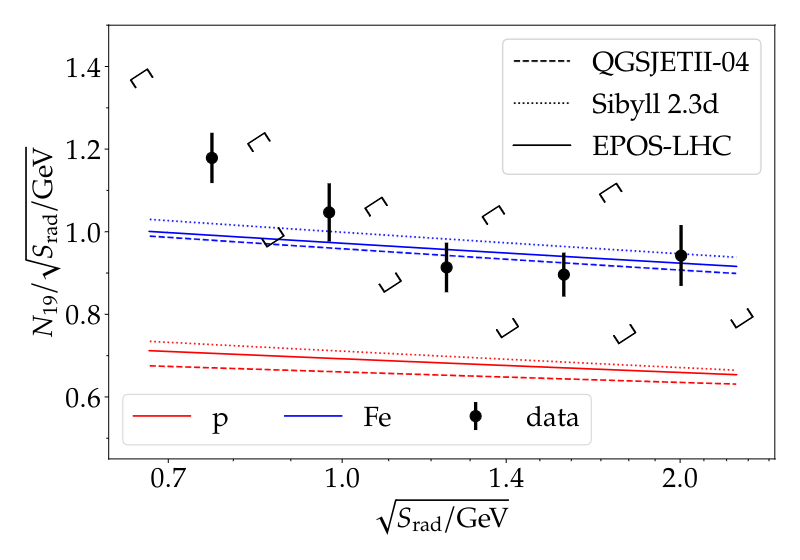}%
    \caption{Histogram showing the reconstruction accuracy of the corrected radiation energy for the subset of high-quality simulated events (left). The inset visualizes mean and standard deviation for different primaries. Normalized muon content as a function of energy estimator (right). The predictions for different hadronic interaction models are denoted by the colored lines for protons and iron primaries. Square brackets indicate the systematic uncertainty of the measurement, the diagonal offsets represent the correlated effect of systematic shifts in the energy estimator.}
    \label{fig:muon_content}
\end{figure}

To validate the LDF model for AERA, a set of more than 1000 air showers is simulated with CoREAS using QGSJet II-04~\cite{PhysRevD.83.014018} as hadronic interaction model and proton and iron nuclei as primary particles. The simulations are reconstructed with a realistic detector simulation adding measured environmental noise from randomly selected timestamps. After applying a selection of high-quality events, we observe a mean underestimation of \SI{3}{\percent} for protons and \SI{1}{\percent} for iron primaries, with a spread of \SI{11}{\percent} in both cases as shown in Fig. \ref{fig:muon_content} (left). Thus, the primary-dependent bias is considered negligible. The remaining bias is likely attributed to signal processing, such as the removal of radio frequency interference. Since this bias is small, it will not be further investigated here but will be accounted for as a systematic uncertainty.

Applying the same selection to approximately ten years of measured data, we obtain 40 hybrid events with energies between \SI{3.4(7)}{EeV} and \SI{12.6(12)}{EeV} as reconstructed by the SD. The most restrictive cut is a threshold of \SI{4}{EeV} for \Eem as reconstructed by AERA based on the energy calibration of Ref~\cite{Schlueter_2023} to ensure full efficiency with the SD. Therefore, this study serves as a proof-of-concept, demonstrating the feasibility of the proposed measurement technique. For the interpretation of the measurements, we obtain predictions of the expected muon number from simulations. We utilize over \num{100,000} inclined air showers simulated with CORSIKA~\cite{corsika} using QGSJet II-04, EPOS-LHC~\cite{PhysRevC.92.034906}, and Sibyll 2.3d~\cite{PhysRevD.102.063002} as high-energy hadronic interaction models using protons and iron nuclei as primaries. While the average muon content normalized by the energy estimator is compatible with the prediction for iron nuclei a lighter composition would also be compatible given the current systematic uncertainties. The presented result is in broad agreement with previous Auger analyses in which a muon deficit in simulations was reported.

\section{Galactic Calibration and Long-Term Stability}

A key requirement for precision measurements with radio detectors is a well-understood and stable calibration over long timescales. Unlike optical techniques, radio detectors have the potential advantage of long-term stability due to the absence of consumable or degradable components such as photomultiplier tubes or optical coatings. Previously, calibration of the AERA stations was achieved by laboratory measurements of the analog signal chain and simulations as well as drone-based measurements of the antennas’ directional response.

\begin{figure}
    \centering
    \includegraphics[width=.5\linewidth]{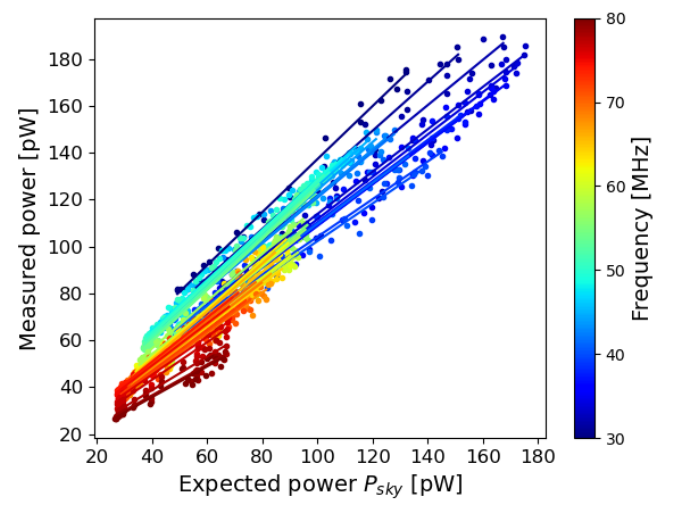}%
    \includegraphics[width=.5\linewidth]{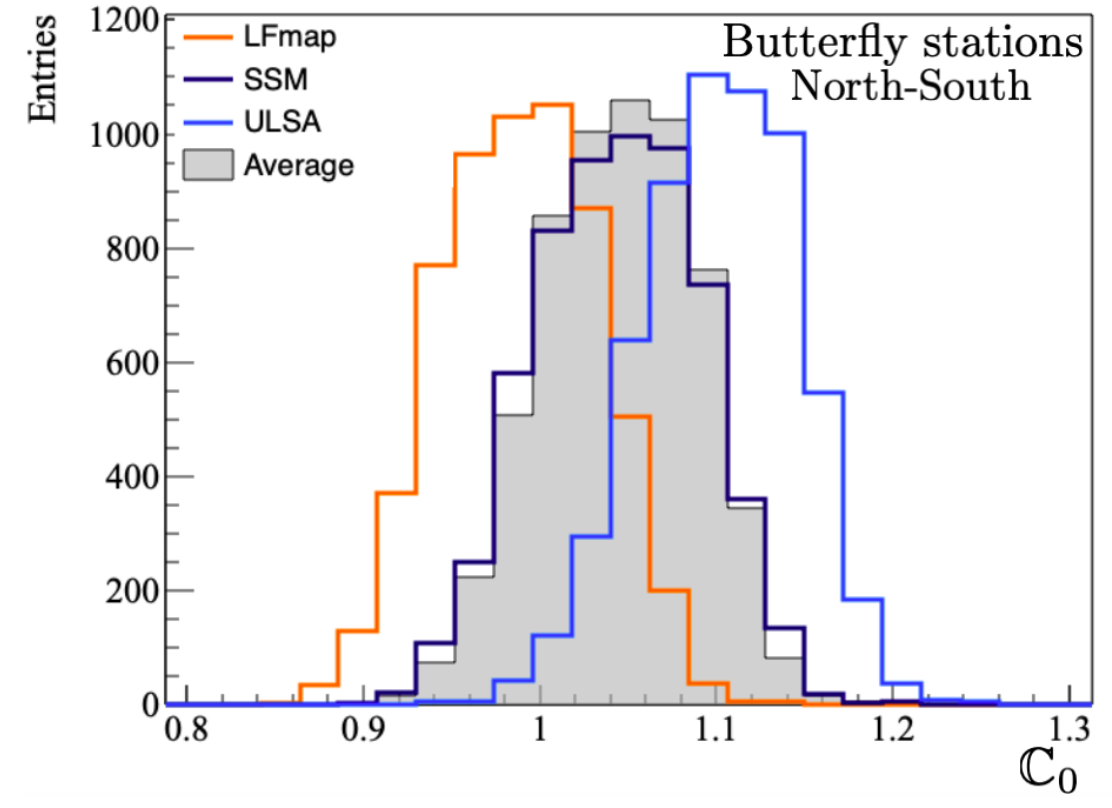}%
    \caption{Measured power versus the expected power based on the LFmap model, as well as the resulting linear fits obtained for each frequency bin (left). These results were determined for the north-south channel of one specific Butterfly antenna using data from January 2019. Distribution of average calibration constants $C_0$ for all Butterfly stations (right), obtained from monthly analysis for several sky models.}
    \label{fig:calibration}
\end{figure}

We present a new calibration method based on Galactic emission. The sidereal modulation of the Galactic signal, driven by the Earth’s rotation, provides a well-characterized and repeatable pattern in the data. Predictions of the received power as a function of local sidereal time and frequency are derived from established sky models of Galactic radio emission. A detailed comparison and discussion of these models has been presented in Ref~\cite{sky_models}. 
These predictions are then compared to the measured power obtained from periodic triggers recorded continuously during data taking. Artifacts such as anthropogenic radio-frequency interference and accidental air-shower signals are identified and removed from the data and the cleaned measurements, corrected for the antenna response, are compared to model predictions using a linear fit in \SI{1}{MHz}-wide frequency bins spanning \SIrange{30}{80}{MHz}.
This yields a set of calibration factors and noise offsets, determined separately for each station, polarization, and frequency bin, allowing for an frequency-dependent calibration. An example of  fits for the 51 frequency bins for a single antenna is shown in Fig.~\ref{fig:calibration} (left).

To estimate the impact of the calibration constants on the cosmic-ray energy uncertainty, we consider a single calibration constant averaged over frequency bins. The results, shown in Fig.~\ref{fig:calibration} (right), are presented for different sky models. Among these, LFmap~\cite{LFmap} and ULSA~\cite{ULSA} approximately represent the lowest and highest values of the calibration constants, respectively. Averaging all models for the full dataset, we obtain a mean calibration constant of $1.04 \pm 0.06$ for the north-south channel of all Butterfly stations. The same procedure yields $1.08 \pm 0.05$ for the east–west channel and $1.01 \pm 0.06$ for both channels of the LPDA stations. The resulst are in general consistent with unity and confirm the results from laboratory-based calibration methods. Interpreting the spread among the models as a systematic uncertainty, we obtain an estimate of approximately \SI{6}{\percent}.

\begin{figure}
    \centering
    \includegraphics[width=\linewidth]{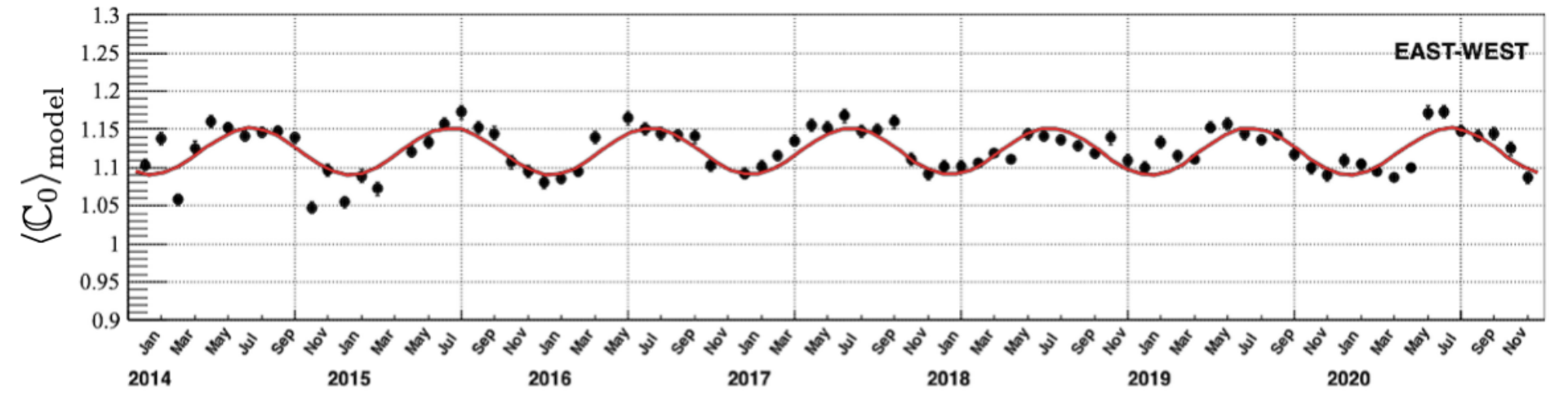}
    \caption{Average calibration constant per month for a single AERA station (black markers) for its east-west-aligned antenna. A cosine fit with an additional linear slope parameter is fitted (red line).}
    \label{fig:long_term_stability}
\end{figure}

In Figure~\ref{fig:long_term_stability}, we show the results of this calibration procedure for a single AERA station. The calibration constants are derived on a monthly basis, enabling a detailed time-resolved study of the detector response. The station’s average calibration constant is slightly above 1.08, reflecting normal station-to-station variations. Over a period of seven years, no significant long-term trend is observed. The seasonal modulation is an understood method artifact due to a varying noise background. By fitting the time evolution of the calibration constants for each station and accounting for uncertainties from both the method and the choice of sky model, we constrain the ageing of the radio-based cosmic-ray energy scale to \SI{-0.27(53)}{\percent}  per decade. 

The results confirm the long term stability and the absence of significant aging effects, highlighting the robustness of radio detection for long-term cosmic-ray observations. Moreover, it supports its use as a reference for cross-calibrating other detector systems, for reducing systematic uncertainties,
and as a tool to monitor potential degradation in other detector components in hybrid systems. 

\section{Conclusion}
We have presented the recent results of the Auger Engineering Radio Array, highlighting its contributions to the measurement of mass composition and muon content. The stability of radio signal in AERA over nearly a decade demonstrates its potential as a calibration standard, potentially reducing the systematic uncertainties in energy scale in the future. These results highlight the value of radio detection as a complementary tool to measure and understand ultra-high-energy cosmic rays. AERA also serves as a platform for developing and validating new analysis techniques, many of which are directly applicable to the AugerPrime Radio Detector in the future.

\clearpage

\section*{The Pierre Auger Collaboration}

{\footnotesize\setlength{\baselineskip}{10pt}
\noindent
\begin{wrapfigure}[11]{l}{0.12\linewidth}
\vspace{-4pt}
\includegraphics[width=0.98\linewidth]{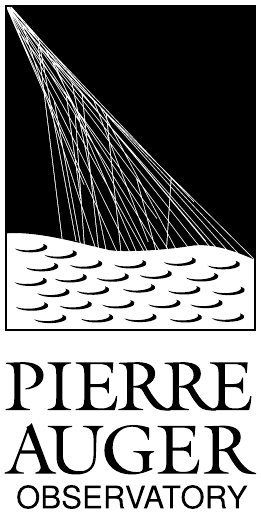}
\end{wrapfigure}
\begin{sloppypar}\noindent
A.~Abdul Halim$^{13}$,
P.~Abreu$^{70}$,
M.~Aglietta$^{53,51}$,
I.~Allekotte$^{1}$,
K.~Almeida Cheminant$^{78,77}$,
A.~Almela$^{7,12}$,
R.~Aloisio$^{44,45}$,
J.~Alvarez-Mu\~niz$^{76}$,
A.~Ambrosone$^{44}$,
J.~Ammerman Yebra$^{76}$
% ...

\end{sloppypar}
\begin{center}
\end{center}

\vspace{1ex}
\begin{description}[labelsep=0.2em,align=right,labelwidth=0.7em,labelindent=0em,leftmargin=2em,noitemsep,before={\renewcommand\makelabel[1]{##1 }}]
\item[$^{1}$] Centro At\'omico Bariloche and Instituto Balseiro (CNEA-UNCuyo-CONICET), San Carlos de Bariloche, Argentina
\item[$^{2}$] Departamento de F\'\i{}sica and Departamento de Ciencias de la Atm\'osfera y los Oc\'eanos, FCEyN, Universidad de Buenos Aires and CONICET, Buenos Aires, Argentina
\item[$^{3}$] IFLP, Universidad Nacional de La Plata and CONICET, La Plata, Argentina
\item[$^{4}$] Instituto de Astronom\'\i{}a y F\'\i{}sica del Espacio (IAFE, CONICET-UBA), Buenos Aires, Argentina
\item[$^{5}$] Instituto de F\'\i{}sica de Rosario (IFIR) -- CONICET/U.N.R.\ and Facultad de Ciencias Bioqu\'\i{}micas y Farmac\'euticas U.N.R., Rosario, Argentina
\item[$^{6}$] Instituto de Tecnolog\'\i{}as en Detecci\'on y Astropart\'\i{}culas (CNEA, CONICET, UNSAM), and Universidad Tecnol\'ogica Nacional -- Facultad Regional Mendoza (CONICET/CNEA), Mendoza, Argentina
\item[$^{7}$] Instituto de Tecnolog\'\i{}as en Detecci\'on y Astropart\'\i{}culas (CNEA, CONICET, UNSAM), Buenos Aires, Argentina
\item[$^{8}$] International Center of Advanced Studies and Instituto de Ciencias F\'\i{}sicas, ECyT-UNSAM and CONICET, Campus Miguelete -- San Mart\'\i{}n, Buenos Aires, Argentina
\item[$^{9}$] Laboratorio Atm\'osfera -- Departamento de Investigaciones en L\'aseres y sus Aplicaciones -- UNIDEF (CITEDEF-CONICET), Argentina
\item[$^{10}$] Observatorio Pierre Auger, Malarg\"ue, Argentina
\item[$^{11}$] Observatorio Pierre Auger and Comisi\'on Nacional de Energ\'\i{}a At\'omica, Malarg\"ue, Argentina
\item[$^{12}$] Universidad Tecnol\'ogica Nacional -- Facultad Regional Buenos Aires, Buenos Aires, Argentina
\item[$^{13}$] University of Adelaide, Adelaide, S.A., Australia
\item[$^{14}$] Universit\'e Libre de Bruxelles (ULB), Brussels, Belgium
\item[$^{15}$] Vrije Universiteit Brussels, Brussels, Belgium
\item[$^{16}$] Centro Brasileiro de Pesquisas Fisicas, Rio de Janeiro, RJ, Brazil
\item[$^{17}$] Centro Federal de Educa\c{c}\~ao Tecnol\'ogica Celso Suckow da Fonseca, Petropolis, Brazil
\item[$^{18}$] Instituto Federal de Educa\c{c}\~ao, Ci\^encia e Tecnologia do Rio de Janeiro (IFRJ), Brazil
\item[$^{19}$] Universidade de S\~ao Paulo, Escola de Engenharia de Lorena, Lorena, SP, Brazil
\item[$^{20}$] Universidade de S\~ao Paulo, Instituto de F\'\i{}sica de S\~ao Carlos, S\~ao Carlos, SP, Brazil
\item[$^{21}$] Universidade de S\~ao Paulo, Instituto de F\'\i{}sica, S\~ao Paulo, SP, Brazil
\item[$^{22}$] Universidade Estadual de Campinas (UNICAMP), IFGW, Campinas, SP, Brazil
\item[$^{23}$] Universidade Estadual de Feira de Santana, Feira de Santana, Brazil
\item[$^{24}$] Universidade Federal de Campina Grande, Centro de Ciencias e Tecnologia, Campina Grande, Brazil
\item[$^{25}$] Universidade Federal do ABC, Santo Andr\'e, SP, Brazil
\item[$^{26}$] Universidade Federal do Paran\'a, Setor Palotina, Palotina, Brazil
\item[$^{27}$] Universidade Federal do Rio de Janeiro, Instituto de F\'\i{}sica, Rio de Janeiro, RJ, Brazil
\item[$^{28}$] Universidad de Medell\'\i{}n, Medell\'\i{}n, Colombia
\item[$^{29}$] Universidad Industrial de Santander, Bucaramanga, Colombia
\item[$^{30}$] Charles University, Faculty of Mathematics and Physics, Institute of Particle and Nuclear Physics, Prague, Czech Republic
\item[$^{31}$] Institute of Physics of the Czech Academy of Sciences, Prague, Czech Republic
\item[$^{32}$] Palacky University, Olomouc, Czech Republic
\item[$^{33}$] CNRS/IN2P3, IJCLab, Universit\'e Paris-Saclay, Orsay, France
\item[$^{34}$] Laboratoire de Physique Nucl\'eaire et de Hautes Energies (LPNHE), Sorbonne Universit\'e, Universit\'e de Paris, CNRS-IN2P3, Paris, France
\item[$^{35}$] Univ.\ Grenoble Alpes, CNRS, Grenoble Institute of Engineering Univ.\ Grenoble Alpes, LPSC-IN2P3, 38000 Grenoble, France
\item[$^{36}$] Universit\'e Paris-Saclay, CNRS/IN2P3, IJCLab, Orsay, France
\item[$^{37}$] Bergische Universit\"at Wuppertal, Department of Physics, Wuppertal, Germany
\item[$^{38}$] Karlsruhe Institute of Technology (KIT), Institute for Experimental Particle Physics, Karlsruhe, Germany
\item[$^{39}$] Karlsruhe Institute of Technology (KIT), Institut f\"ur Prozessdatenverarbeitung und Elektronik, Karlsruhe, Germany
\item[$^{40}$] Karlsruhe Institute of Technology (KIT), Institute for Astroparticle Physics, Karlsruhe, Germany
\item[$^{41}$] RWTH Aachen University, III.\ Physikalisches Institut A, Aachen, Germany
\item[$^{42}$] Universit\"at Hamburg, II.\ Institut f\"ur Theoretische Physik, Hamburg, Germany
\item[$^{43}$] Universit\"at Siegen, Department Physik -- Experimentelle Teilchenphysik, Siegen, Germany
\item[$^{44}$] Gran Sasso Science Institute, L'Aquila, Italy
\item[$^{45}$] INFN Laboratori Nazionali del Gran Sasso, Assergi (L'Aquila), Italy
\item[$^{46}$] INFN, Sezione di Catania, Catania, Italy
\item[$^{47}$] INFN, Sezione di Lecce, Lecce, Italy
\item[$^{48}$] INFN, Sezione di Milano, Milano, Italy
\item[$^{49}$] INFN, Sezione di Napoli, Napoli, Italy
\item[$^{50}$] INFN, Sezione di Roma ``Tor Vergata'', Roma, Italy
\item[$^{51}$] INFN, Sezione di Torino, Torino, Italy
\item[$^{52}$] Istituto di Astrofisica Spaziale e Fisica Cosmica di Palermo (INAF), Palermo, Italy
\item[$^{53}$] Osservatorio Astrofisico di Torino (INAF), Torino, Italy
\item[$^{54}$] Politecnico di Milano, Dipartimento di Scienze e Tecnologie Aerospaziali , Milano, Italy
\item[$^{55}$] Universit\`a del Salento, Dipartimento di Matematica e Fisica ``E.\ De Giorgi'', Lecce, Italy
\item[$^{56}$] Universit\`a dell'Aquila, Dipartimento di Scienze Fisiche e Chimiche, L'Aquila, Italy
\item[$^{57}$] Universit\`a di Catania, Dipartimento di Fisica e Astronomia ``Ettore Majorana``, Catania, Italy
\item[$^{58}$] Universit\`a di Milano, Dipartimento di Fisica, Milano, Italy
\item[$^{59}$] Universit\`a di Napoli ``Federico II'', Dipartimento di Fisica ``Ettore Pancini'', Napoli, Italy
\item[$^{60}$] Universit\`a di Palermo, Dipartimento di Fisica e Chimica ''E.\ Segr\`e'', Palermo, Italy
\item[$^{61}$] Universit\`a di Roma ``Tor Vergata'', Dipartimento di Fisica, Roma, Italy
\item[$^{62}$] Universit\`a Torino, Dipartimento di Fisica, Torino, Italy
\item[$^{63}$] Benem\'erita Universidad Aut\'onoma de Puebla, Puebla, M\'exico
\item[$^{64}$] Unidad Profesional Interdisciplinaria en Ingenier\'\i{}a y Tecnolog\'\i{}as Avanzadas del Instituto Polit\'ecnico Nacional (UPIITA-IPN), M\'exico, D.F., M\'exico
\item[$^{65}$] Universidad Aut\'onoma de Chiapas, Tuxtla Guti\'errez, Chiapas, M\'exico
\item[$^{66}$] Universidad Michoacana de San Nicol\'as de Hidalgo, Morelia, Michoac\'an, M\'exico
\item[$^{67}$] Universidad Nacional Aut\'onoma de M\'exico, M\'exico, D.F., M\'exico
\item[$^{68}$] Institute of Nuclear Physics PAN, Krakow, Poland
\item[$^{69}$] University of \L{}\'od\'z, Faculty of High-Energy Astrophysics,\L{}\'od\'z, Poland
\item[$^{70}$] Laborat\'orio de Instrumenta\c{c}\~ao e F\'\i{}sica Experimental de Part\'\i{}culas -- LIP and Instituto Superior T\'ecnico -- IST, Universidade de Lisboa -- UL, Lisboa, Portugal
\item[$^{71}$] ``Horia Hulubei'' National Institute for Physics and Nuclear Engineering, Bucharest-Magurele, Romania
\item[$^{72}$] Institute of Space Science, Bucharest-Magurele, Romania
\item[$^{73}$] Center for Astrophysics and Cosmology (CAC), University of Nova Gorica, Nova Gorica, Slovenia
\item[$^{74}$] Experimental Particle Physics Department, J.\ Stefan Institute, Ljubljana, Slovenia
\item[$^{75}$] Universidad de Granada and C.A.F.P.E., Granada, Spain
\item[$^{76}$] Instituto Galego de F\'\i{}sica de Altas Enerx\'\i{}as (IGFAE), Universidade de Santiago de Compostela, Santiago de Compostela, Spain
\item[$^{77}$] IMAPP, Radboud University Nijmegen, Nijmegen, The Netherlands
\item[$^{78}$] Nationaal Instituut voor Kernfysica en Hoge Energie Fysica (NIKHEF), Science Park, Amsterdam, The Netherlands
\item[$^{79}$] Stichting Astronomisch Onderzoek in Nederland (ASTRON), Dwingeloo, The Netherlands
\item[$^{80}$] Universiteit van Amsterdam, Faculty of Science, Amsterdam, The Netherlands
\item[$^{81}$] Case Western Reserve University, Cleveland, OH, USA
\item[$^{82}$] Colorado School of Mines, Golden, CO, USA
\item[$^{83}$] Department of Physics and Astronomy, Lehman College, City University of New York, Bronx, NY, USA
\item[$^{84}$] Michigan Technological University, Houghton, MI, USA
\item[$^{85}$] New York University, New York, NY, USA
\item[$^{86}$] University of Chicago, Enrico Fermi Institute, Chicago, IL, USA
\item[$^{87}$] University of Delaware, Department of Physics and Astronomy, Bartol Research Institute, Newark, DE, USA
\item[] -----
\item[$^{a}$] Max-Planck-Institut f\"ur Radioastronomie, Bonn, Germany
\item[$^{b}$] also at Kapteyn Institute, University of Groningen, Groningen, The Netherlands
\item[$^{c}$] School of Physics and Astronomy, University of Leeds, Leeds, United Kingdom
\item[$^{d}$] Fermi National Accelerator Laboratory, Fermilab, Batavia, IL, USA
\item[$^{e}$] Pennsylvania State University, University Park, PA, USA
\item[$^{f}$] Colorado State University, Fort Collins, CO, USA
\item[$^{g}$] Louisiana State University, Baton Rouge, LA, USA
\item[$^{h}$] now at Graduate School of Science, Osaka Metropolitan University, Osaka, Japan
\item[$^{i}$] Institut universitaire de France (IUF), France
\item[$^{j}$] now at Technische Universit\"at Dortmund and Ruhr-Universit\"at Bochum, Dortmund and Bochum, Germany
\end{description}

\section*{Acknowledgments}

\begin{sloppypar}
The successful installation, commissioning, and operation of the Pierre
Auger Observatory would not have been possible without the strong
commitment and effort from the technical and administrative staff in
Malarg\"ue. We are very grateful to the following agencies and
organizations for financial support:
\end{sloppypar}

\begin{sloppypar}
Argentina -- Comisi\'on Nacional de Energ\'\i{}a At\'omica; Agencia Nacional de
Promoci\'on Cient\'\i{}fica y Tecnol\'ogica (ANPCyT); Consejo Nacional de
Investigaciones Cient\'\i{}ficas y T\'ecnicas (CONICET); Gobierno de la
Provincia de Mendoza; Municipalidad de Malarg\"ue; NDM Holdings and Valle
Las Le\~nas; in gratitude for their continuing cooperation over land
access; Australia -- the Australian Research Council; Belgium -- Fonds
de la Recherche Scientifique (FNRS); Research Foundation Flanders (FWO),
Marie Curie Action of the European Union Grant No.~101107047; Brazil --
Conselho Nacional de Desenvolvimento Cient\'\i{}fico e Tecnol\'ogico (CNPq);
Financiadora de Estudos e Projetos (FINEP); Funda\c{c}\~ao de Amparo \`a
Pesquisa do Estado de Rio de Janeiro (FAPERJ); S\~ao Paulo Research
Foundation (FAPESP) Grants No.~2019/10151-2, No.~2010/07359-6 and
No.~1999/05404-3; Minist\'erio da Ci\^encia, Tecnologia, Inova\c{c}\~oes e
Comunica\c{c}\~oes (MCTIC); Czech Republic -- GACR 24-13049S, CAS LQ100102401,
MEYS LM2023032, CZ.02.1.01/0.0/0.0/16{\textunderscore}013/0001402,
CZ.02.1.01/0.0/0.0/18{\textunderscore}046/0016010 and
CZ.02.1.01/0.0/0.0/17{\textunderscore}049/0008422 and CZ.02.01.01/00/22{\textunderscore}008/0004632;
France -- Centre de Calcul IN2P3/CNRS; Centre National de la Recherche
Scientifique (CNRS); Conseil R\'egional Ile-de-France; D\'epartement
Physique Nucl\'eaire et Corpusculaire (PNC-IN2P3/CNRS); D\'epartement
Sciences de l'Univers (SDU-INSU/CNRS); Institut Lagrange de Paris (ILP)
Grant No.~LABEX ANR-10-LABX-63 within the Investissements d'Avenir
Programme Grant No.~ANR-11-IDEX-0004-02; Germany -- Bundesministerium
f\"ur Bildung und Forschung (BMBF); Deutsche Forschungsgemeinschaft (DFG);
Finanzministerium Baden-W\"urttemberg; Helmholtz Alliance for
Astroparticle Physics (HAP); Helmholtz-Gemeinschaft Deutscher
Forschungszentren (HGF); Ministerium f\"ur Kultur und Wissenschaft des
Landes Nordrhein-Westfalen; Ministerium f\"ur Wissenschaft, Forschung und
Kunst des Landes Baden-W\"urttemberg; Italy -- Istituto Nazionale di
Fisica Nucleare (INFN); Istituto Nazionale di Astrofisica (INAF);
Ministero dell'Universit\`a e della Ricerca (MUR); CETEMPS Center of
Excellence; Ministero degli Affari Esteri (MAE), ICSC Centro Nazionale
di Ricerca in High Performance Computing, Big Data and Quantum
Computing, funded by European Union NextGenerationEU, reference code
CN{\textunderscore}00000013; M\'exico -- Consejo Nacional de Ciencia y Tecnolog\'\i{}a
(CONACYT) No.~167733; Universidad Nacional Aut\'onoma de M\'exico (UNAM);
PAPIIT DGAPA-UNAM; The Netherlands -- Ministry of Education, Culture and
Science; Netherlands Organisation for Scientific Research (NWO); Dutch
national e-infrastructure with the support of SURF Cooperative; Poland
-- Ministry of Education and Science, grants No.~DIR/WK/2018/11 and
2022/WK/12; National Science Centre, grants No.~2016/22/M/ST9/00198,
2016/23/B/ST9/01635, 2020/39/B/ST9/01398, and 2022/45/B/ST9/02163;
Portugal -- Portuguese national funds and FEDER funds within Programa
Operacional Factores de Competitividade through Funda\c{c}\~ao para a Ci\^encia
e a Tecnologia (COMPETE); Romania -- Ministry of Research, Innovation
and Digitization, CNCS-UEFISCDI, contract no.~30N/2023 under Romanian
National Core Program LAPLAS VII, grant no.~PN 23 21 01 02 and project
number PN-III-P1-1.1-TE-2021-0924/TE57/2022, within PNCDI III; Slovenia
-- Slovenian Research Agency, grants P1-0031, P1-0385, I0-0033, N1-0111;
Spain -- Ministerio de Ciencia e Innovaci\'on/Agencia Estatal de
Investigaci\'on (PID2019-105544GB-I00, PID2022-140510NB-I00 and
RYC2019-027017-I), Xunta de Galicia (CIGUS Network of Research Centers,
Consolidaci\'on 2021 GRC GI-2033, ED431C-2021/22 and ED431F-2022/15),
Junta de Andaluc\'\i{}a (SOMM17/6104/UGR and P18-FR-4314), and the European
Union (Marie Sklodowska-Curie 101065027 and ERDF); USA -- Department of
Energy, Contracts No.~DE-AC02-07CH11359, No.~DE-FR02-04ER41300,
No.~DE-FG02-99ER41107 and No.~DE-SC0011689; National Science Foundation,
Grant No.~0450696, and NSF-2013199; The Grainger Foundation; Marie
Curie-IRSES/EPLANET; European Particle Physics Latin American Network;
and UNESCO.
\end{sloppypar}

} 

\end{document}